\documentclass{arxiv}
\usepackage{epsfig}
\usepackage{subfig}
\usepackage{cite}
\begin{document}

\catchline{}{}{}{}{}

\title{The leading digit distribution of the worldwide Illicit Financial Flows
}

\author{T. A. Mir$^{*}$}

\address{Nuclear Research Laboratory, Astrophysical Sciences Division,\\ Bhabha Atomic Research Centre,\\  Srinagar-190 006, Jammu and Kashmir, India\\
$^{*}$taarik.mir@gmail.com}

\maketitle


\section{Abstract}

Benford's law states that in data sets from different phenomena leading digits tend to be distributed logarithmically such that the numbers beginning with smaller digits occur more often than those with larger ones. Particularly, the law is known to hold for different types of financial data.  The Illicit Financial Flows (IFFs) exiting the developing countries are frequently discussed as hidden resources which could have been otherwise properly utilized for their development. We investigate here the distribution of the leading digits in the recent data on estimates of IFFs to look for the existence of a pattern as predicted by Benford's law and establish that the frequency of occurrence of the leading digits in these estimates does closely follow the law.
\section{Keywords}

Benford's law; developing countries; illicit financial flows 


\section{Introduction}	

The leading digit phenomenon stands for a counter-intuitive observation first made by S. Newcomb who found the starting pages of logarithmic table books to be dirtier than the last ones and attributed this effect to the fact that numbers with smaller first non-zero digits are more often looked for\cite{Newcomb}. The curious observation went quite unnoticed till F. Benford rediscovered it and by analyzing data from diverse fields established the law in the form of following empirical mathematical equation\cite{Benford}
\begin{equation}
P(d)= log_{10}(1+\frac{1}{d}), d= 1, 2, 3...,9
\end{equation}
where $P(d)$ is the probability of a number having the first non-zero digit d and $log_{10}$ is logarithmic to base 10. Thus in any given data set the theoretical probability of occurrence of digits 1 to 9 as the leading digit is as shown in Table 1.
\begin{table}[h]
\tbl{The Benford distribution of leading digits}
{\begin{tabular}{@{}lllllllllll@{}} \toprule
Digit \hphantom{00} & 1 & 2 & 3 & 4 & 5 & 6 & 7 & 8 & 9 \\
Proportion \hphantom{00} & 0.301 & 0.176 & 0.125 & 0.097 & 0.079 & 0.067 & 0.058 & 0.051 & 0.046 \\ \botrule
\end{tabular} \label{ta1}}
\end{table} 
\newline
Though a complete explanation of Benford's law is outstanding, significant advances have been made in the understanding of this ubiquitous law\cite{Berger}. It has been found to be scale invariant, being the only digit law to be so, which means that a change in the units of data measurement does not affect the validity of the law\cite{Pinkham}. This scale-invariance was further shown to imply base-invariance, being independent of the base 2, 8 or 10 in which  numbers are expressed\cite{Hill}.  Further the law has been shown to arise naturally for processes whose time evolutions are governed by multiplicative fluctuations\cite{Pietronero}.
\newline
Data from a variety of phenomena have been found to follow Benford's law. The leading digits of demographic data like populations of cities\cite{Benford}, countries\cite{Sandron} and world religions\cite{Mir} are distributed according to this law. Similarly scientific data like complex atomic spectra\cite{Pain}, full decay widths of hadrons\cite{Shao} and magnitude of earth quakes\cite{Sambridge} follow the law. But interest in the peculiar law surged due to its frequent appearance amongst the various types of financial data. The leading digits of stock market data\cite{Pietronero, Ley}, winning bids for certain eBay auctions\cite{Giles}, health insurance data\cite{Maher} and the invoice data of oil companies\cite{Nigrini1} are distributed as predicted by Benford's law.
\newline
Benford's law has been used to assess the quality of the macroeconomic data submitted by countries to the World Bank and significant deviations from the law suggesting deliberate falsification have been found for the data from the developing countries\cite{Nye, Gonzalez}. Further, deviations from Benford's law support the hypothesis of occasional misreporting of economic data by some countries for strategic purposes\cite{Michalski}. A recent example is the macroeconomic data relevant to the deficit criteria reported by Greece, a member state of European Union, to Eurostat for which greatest deviation from Benford's law was found amongst the data from all euro states\cite{Rauch}.
\newline
In the context of the current financial crisis an overhaul of the whole global financial system is necessitated. Therefore, the estimates of IFFs have generated a lot of media and public interest\cite{tax, Reuter}. Due to their direct relation to corruption and crime, IFFs have also been identified as major impediments to upliftment of social standards across developing countries\cite{UN1}. We investigate here whether the data on IFFs from developing countries exhibits the pattern in the distribution of the leading digits as predicted by Benford's law.

\section{Data}
The data source for the present analysis is the Global Financial Integrity (GFI), a research and advocacy organization working to curtail IFFs out of developing countries\cite{GFI}. Researchers at GFI by the application of the current economic models to the most recent macroeconomic data available, estimate the volume and pattern of IFFs exiting the developing world over different periods of time. We analyze four reports of GFI for checking the conformity of the data to Benford's law\cite{kar, kar1, kar2, kar3}.

\subsection{Data analysis and Results}

Based on the macroeconomic data available from international financial institutions, the GFI reports study the IFFs from 160 developing countries of the world grouped into five regions. The entire list is normalized to exclude the countries for which the illicit flows don't exist by subjecting it to a two-stage filtration process (i) out of the five years period outflows must exist for at least three years and (ii) exceed the threshold (10 percent) with respect to exports\cite{kar1}. The restrictions imposed by the filtration process give conservative or low end estimates of such financial flows from developing countries. Countries that fail to pass through either stages of the filtration process are eliminated from the list. Thus for each report we have a large non-normalized and a slightly smaller normalized list of countries and consequently for each report we analyze both the lists. 
\newline 
We detail the statistical analysis of the IFFs data from the four reports in Tables 2-5. The $N_{Obs}$, the number of times each digit from 1 to 9 (column 1) appears as leading digit are shown in subsequent columns along with $N_{Ben}$ the corresponding frequency as predicted by Benford's law: 
\begin{equation}
N_{Ben}= N log_{10}(1+\frac{1}{d})
\end{equation} 
where $N$ is the total number of countries for which the IFFs data is reported. A visual inspection of the tables shows clearly that the observed digit frequencies closely match those predicted by Benford's law. However, impossibility of having an actual proportion of leading digits that is exactly equal to Benford proportion necessitates a quantification of the degree of closeness of the two\cite{Nigrini3}. To determine an acceptable range for the observed distribution we use three statistical approaches. First to assess whether the observed frequencies of the leading digits are consistent with an acceptable range of the theoretically predicted values we calculate the corresponding confidence intervals (CI) by the following formula\cite{Minteer}
\begin{equation}
E= Z_{critical}\sqrt{P(d)(1-P(d))/N}
\end{equation} 
where $Z_{critical}$ is the Z-value associated with the desired CI, $P(d)$ is defined in equation (1). The CI's are shown in brackets along with $N_{Ben}$, the predicted digit frequencies. For 95\% CI the $Z_{critical}$ is 1.96\cite{Durtschi}. For example, as shown in column 2 of Table 2 out of a total of N=144, the observed count for digit 1 as leading digit is 37 whereas that expected from Benford's law is 43.3 with a confidence interval of $\pm$7.5.
\newline 
Further, to determine the overall behavior of the observed distribution with respect to that predicted by Benford's law we state the \textit{Null Hypothesis}, $H_{O}$ that the observed and predicted distributions of the leading digits are same and then to examine the goodness-of-fit we use Pearson's $\chi^{2}$ test defined as
\begin{equation}
  \chi^{2}(n-1)=\sum_{i=1}^n\dfrac{(N_{Obs}-N_{Ben})^{2}}{N_{Ben}}
\end{equation}

For a data set with $n-1=9-1=8$ degrees of freedom, the critical value of $\chi^{2}$ at 5\% level of significance is $15.507$. If the value of the calculated $\chi^{2}$ is less than the critical value then we accept the null hypothesis and conclude that the data fits Benford's law.
\newline
$\chi^{2}$ test is the primary choice of researchers for examining the goodness-of-fit of a set of data to a Benford distribution. However, rejection of the null hypothesis becomes difficult if the number of observations in a data set is small. Further, for a distribution decreasing like a Benford one, the Kolmogorov-Smirnoff (K-S) goodness-of-fit test has been shown to be more powerful than the traditional $\chi^{2}$\cite{Steele}. The K-S test is based on the cumulative density function and uses the largest of the absolute of the differences between the expected and the observed cumulative proportions of the leading digits. The critical values of the K-S test are given by the equation
\begin{equation}
  K-S_{ critical} =\frac{1.36}{\sqrt{N}}
\end{equation}
where 1.36 is the constant for 5\% level of significance and N is the number of observations\cite{Nigrini3}.
\newline
In Table 2 we summarize the observed distribution of the leading digits for the three data sets from the 2008 report (Tables 18 and 19) of GFI which covers the IFFs data for the period of 2002-2006\cite{kar1}. For each data set, the calculated $\chi^{2}$ and the K-S test statistic are respectively shown in the last two rows of Table 2. As can be seen from column 2 of last but one row, the $\chi^{2}$ of 6.775 for the non-normalized list is less than the critical value of $15.507$. Further, for the same data set the calculated K-S test statistic (the last row and column 2) is 0.074, a value which is again less than the critical value (as shown) of 0.113. Hence both the tests call for the acceptance of the null hypothesis which means that the non-normalized IFFs data closely resembles a Benford distribution. After elimination of the 45 countries via the normalization only 99 countries (column 3) are left in the list for which again both the $\chi^{2}$ of 2.766 and K-S test statistic of 0.048 turn out be far less than the respective critical values of 15.507 and 0.137 and the null hypothesis again is accepted. Furthermore, in column 4 we show the statistics for the non-normalized IFFs data of 119 countries estimated using the World Bank Changes in External Debt (WB CED) model. Both the $\chi^{2}$ of 7.477 and K-S statistic of 0.077 again turn out to be less than the respective cutoff values making the null hypothesis acceptable.  
\begin{table}[]
\tbl{The leading digit distribution of country-wise yearly average non-normalized, yearly average normalized and yearly average non-normalized (Average WB CED) IFFs estimates: 2002-2006 (millions of U.S. dollars)}
{\begin{tabular}{@{}llllll@{}} \toprule
First Digit & (N=144) & (N=99) & (N=119) \\ \colrule
$1$ \hphantom{00} & 37 (43.3$\pm$7.5) & 31 (29.8$\pm$9.0)  & 34 (35.8$\pm$8.2)\\
$2$ \hphantom{00} & 21 (25.4$\pm$6.2) & 21 (17.4$\pm$7.5)  & 19 (21.0$\pm$6.8)\\
$3$ \hphantom{00} & 23 (18.0$\pm$5.4) & 11 (12.4$\pm$6.5)  & 14 (14.9$\pm$5.9)\\
$4$ \hphantom{00} & 13 (14.0$\pm$4.8) & 6 (9.6$\pm$5.8)    & 7 (11.5$\pm$5.3)\\
$5$ \hphantom{00} & 11 (11.4$\pm$4.4) & 9  (7.8$\pm$5.3)   & 14 (9.4$\pm$4.9)\\
$6$ \hphantom{00} & 14 (9.6$\pm$4.1) & 7 (6.6$\pm$4.9)     & 12 (8.0$\pm$4.5)\\
$7$ \hphantom{00} & 10 (8.4$\pm$3.8) & 6  (5.7$\pm$4.6)    & 5 (7.0$\pm$4.2)\\ 
$8$ \hphantom{00} & 10 (7.4$\pm$3.6) & 4  (5.1$\pm$4.3)    & 7 (6.1$\pm$4.0)\\
$9$ \hphantom{00} & 5 (6.7$\pm$3.4)  & 4 (4.5$\pm$4.1)     & 7 (5.4$\pm$3.8) \\
\botrule
\textbf{Pearson} $\chi^{2}$ \hphantom{00} &  \bf6.775 & \bf2.766 & \bf7.477 \\ \botrule
\textbf{K-S test}  \hphantom{00} & \bf0.074$<$0.113 &  \bf0.048$<$0.137 & \bf0.077$<$0.125\\ \botrule
\end{tabular} \label{ta1}}
\end{table} 
\begin{figure}
\begin{minipage}[b]{.9\linewidth}
\vspace*{-25pt}
\hspace*{20pt}
\centering
\begin{tabular}{cc}
\hspace*{-80pt}
\vspace*{-90pt}
\epsfig{file=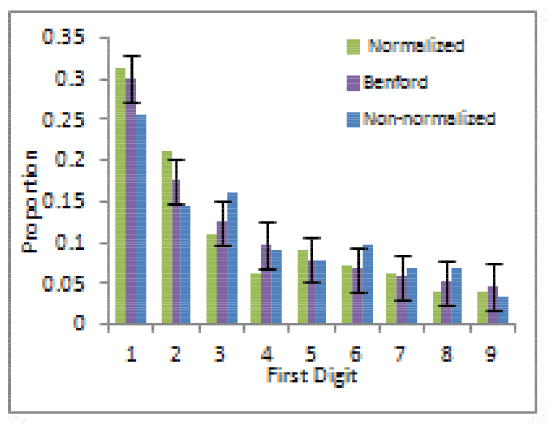,width=0.7\linewidth, height=0.7\linewidth, angle=270,clip=} &
\hspace*{-60pt}
\epsfig{file=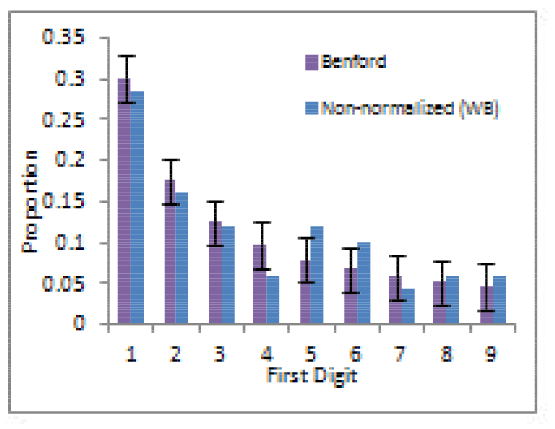,width=0.7\linewidth, height=0.7\linewidth, angle=270, clip=}\\
\end{tabular}
\vspace*{100pt}
\end{minipage}
\hspace*{80pt}
\vspace*{-30pt}
\caption{Observed and Benford distributions of leading digits for country-wise yearly average normalized, yearly average non-normalized and yearly average non-normalized (Average WB CED) IFFs: 2002-2006 (millions of U.S. dollars)}
\end{figure}

\begin{figure}
\begin{minipage}[b]{.9\linewidth}
\vspace*{-25pt}
\hspace*{-10pt}
\centering
\begin{tabular}{cc}
\hspace*{-50pt}
\vspace*{10pt}
\epsfig{file=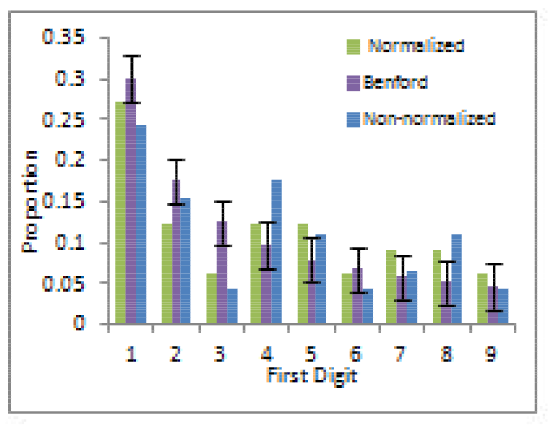,width=0.7\linewidth, height=0.7\linewidth, angle=270,clip=} &
\hspace*{-60pt}
\epsfig{file=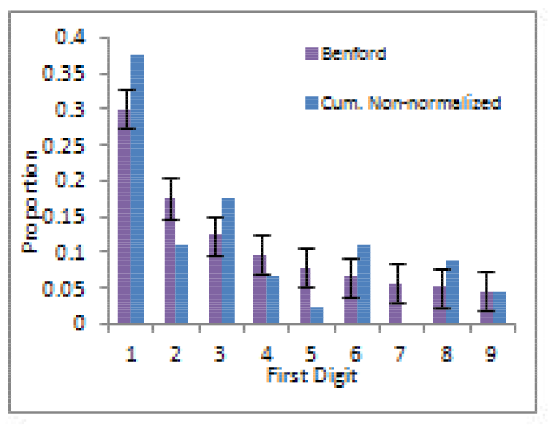,width=0.7\linewidth, height=0.7\linewidth, angle=270, clip=}\\
\end{tabular}
\vspace*{-20pt}
\end{minipage}
\caption{Observed and Benford distributions of leading digits of yearly average normalized, yearly average non-normalized and  cumulative non-normalized IFFs estimates for 1990-2008 for least developed countries (millions of U.S. dollars)}
\end{figure}

\begin{table}[]
\tbl{The leading digit distribution of  yearly average non-normalized, yearly average normalized and cumulative non-normalized IFFs estimates: 1990-2008 for least developed countries (millions of U.S. dollars)}
{\begin{tabular}{@{}llllll@{}} \toprule
First Digit & (N=45) & (N=33) & (N=45)   \\ \colrule
$1$ \hphantom{00} & 11 (13.6$\pm$3.1)  & 9 (9.9$\pm$2.6) &  17 (13.5$\pm$3.1)\\
$2$ \hphantom{00} & 7 (7.9$\pm$2.6) & 4 (5.8$\pm$2.2) & 5 (7.9$\pm$2.6) \\
$3$ \hphantom{00} & 2 (5.6$\pm$2.2) & 2 (4.1$\pm$1.9) & 8 (5.6$\pm$2.2) \\
$4$ \hphantom{00} & 8 (4.4$\pm$2.0) & 4 (3.2$\pm$1.7) &  3 (4.4$\pm$2.0)\\ 
$5$ \hphantom{00} & 5 (3.6$\pm$1.8) & 4 (2.6$\pm$1.6) &  1 (3.6$\pm$1.8) \\
$6$ \hphantom{00} & 2 (3.0$\pm$1.7)  & 2 (2.2$\pm$1.4) &  5 (3.0$\pm$1.7)\\
$7$ \hphantom{00} & 3 (2.6$\pm$1.6) & 3 (1.9$\pm$1.3) &  0 (2.6$\pm$1.6) \\
$8$ \hphantom{00} & 5 (2.3$\pm$1.5)  & 3 (1.7$\pm$1.3) & 4 (2.3$\pm$1.5) \\
$9$ \hphantom{00} & 2 (2.1$\pm$1.4)  & 2 (1.5$\pm$1.2) &  2 (2.1$\pm$1.4)\\ 
\botrule
\textbf{Pearson} $\chi^{2}$ \hphantom{00} & \bf10.110  & \bf4.498 & \bf10.408 \\ \botrule
\textbf{K-S test}  \hphantom{00} & \bf0.158$<$0.203 & \bf0.148$<$0.237 &  \bf0.077$<$0.203 \\ \botrule
\end{tabular} \label{ta1}}
\end{table}

\begin{table}[]
\tbl{The leading digit distribution of country-wise largest average non-normalized, largest average normalized , cumulative non-normalized, cumulative normalized IFFs estimates: 2000-2008 (millions of U.S. dollars)}
{\begin{tabular}{@{}llllll@{}} \toprule
First Digit 
& (N=152) & (N=125) & (N=154) & (N=127) \\ \colrule
$1$ \hphantom{00} & 41 (45.6$\pm$7.3) & 40 (37.6$\pm$8.0) &  44 (46.4$\pm$7.3) & 43 (38.2$\pm$8.0) \\
$2$ \hphantom{00} & 27 (26.8$\pm$6.1) & 21 (22.0$\pm$6.7) &  28 (27.1$\pm$6.0) & 19 (22.4$\pm$6.6) \\
$3$ \hphantom{00} & 20 (19.0$\pm$5.3) & 18 (15.6$\pm$5.8) &  20 (19.2$\pm$5.2) & 21 (15.9$\pm$5.8) \\
$4$ \hphantom{00} & 15 (14.7$\pm$4.7) & 13 (12.1$\pm$5.2) &  18 (14.9$\pm$4.7) & 10 (12.3$\pm$5.2) \\ 
$5$ \hphantom{00} & 19 (12.0$\pm$4.3) & 12 (9.9$\pm$4.7)  &  15 (12.2$\pm$4.3) & 13 (10.1$\pm$4.7) \\
$6$ \hphantom{00} & 10 (10.2$\pm$4.0) & 6 (8.4$\pm$4.4)   &  11 (10.3$\pm$4.0) & 8 (8.5$\pm$4.4) \\
$7$ \hphantom{00} & 10 (8.8$\pm$3.7)  & 8 (7.2$\pm$4.1)   &  4 (8.9$\pm$3.7) & 4 (7.4$\pm$4.1) \\
$8$ \hphantom{00} & 6 (7.8$\pm$3.5)   & 4  (6.4$\pm$3.9)  &  6 (7.9$\pm$3.5) & 4 (6.5$\pm$3.8) \\
$9$ \hphantom{00} & 4 (7.0$\pm$3.3)   & 3 (5.7$\pm$3.7)   &  8  (7.0$\pm$3.3) & 5 (5.8$\pm$3.6)\\ 
\botrule
\textbf{Pearson} $\chi^{2}$ \hphantom{00} & \bf6.409 &  \bf4.008& \bf4.803  & \bf6.695 \\ \botrule
\textbf{K-S test}  \hphantom{00} & \bf0.031$<$0.110 &  \bf0.054$<$0.122 & \bf0.038$<$0.110  & \bf0.056$<$0.121 \\ \botrule
\end{tabular} \label{ta2}}
\end{table}
\begin{figure}
\begin{minipage}[b]{.9\linewidth}
\vspace*{-25pt}
\hspace*{-10pt}
\centering
\begin{tabular}{cc}
\hspace*{-50pt}
\vspace*{-95pt}
\epsfig{file=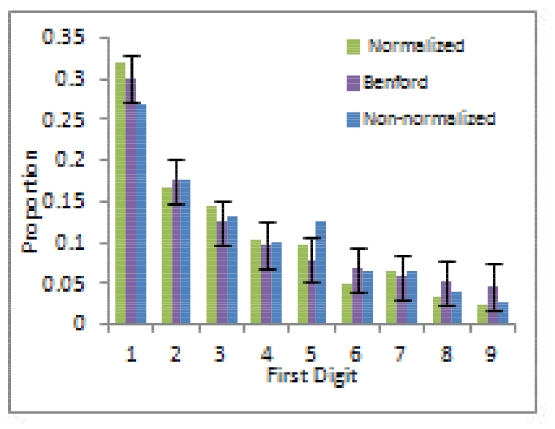,width=0.7\linewidth, height=0.7\linewidth, angle=270,clip=} &
\hspace*{-60pt}
\epsfig{file=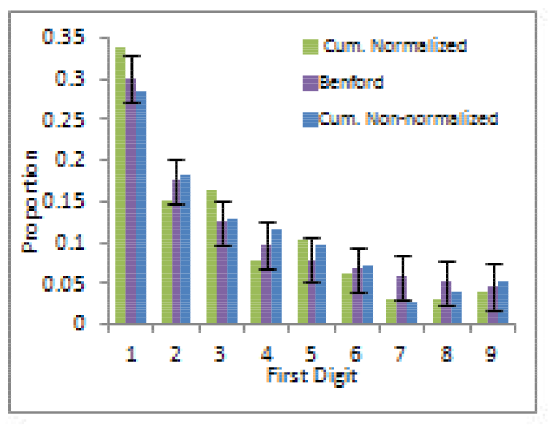,width=0.7\linewidth, height=0.7\linewidth, angle=270, clip=}\\
\vspace*{40pt}
\hspace*{-130pt}
\end{tabular}
\vspace*{35pt}
\hspace*{20pt}
\end{minipage}
\caption{Observed and Benford distributions of leading digits for country-wise largest average normalized, largest average non-normalized, cumulative normalized and cumulative non-normalized IFFs: 2000-2008 (millions of U.S. dollars)}
\end{figure}

\begin{table}[]
\tbl{The leading digit distribution of country-wise largest average non-normalized, largest average normalized, cumulative non-normalized, cumulative normalized IFFs estimates: 2000-2009 (millions of U.S. dollars)}
{\begin{tabular}{@{}llllll@{}} \toprule
First Digit & (N=157) & (N=114) & (N=157) & (N=116) \\ \colrule
$1$ \hphantom{00} & 41 (47.3$\pm$7.2) & 37 (34.3$\pm$8.4) &  45 (47.3$\pm$7.2) & 36 (34.9$\pm$8.4) \\
$2$ \hphantom{00} & 29 (27.7$\pm$6.0) & 12 (20.1$\pm$7.0) &  27 (27.7$\pm$6.0) & 16 (20.4$\pm$6.9)\\
$3$ \hphantom{00} & 17 (19.6$\pm$5.2) & 16 (14.2$\pm$6.1) &  15 (19.6$\pm$5.2) & 15 (14.5$\pm$6.0) \\
$4$ \hphantom{00} & 22 (15.2$\pm$4.6) & 15 (11.0$\pm$5.4) &  22 (15.2$\pm$4.6) & 15 (11.2$\pm$5.4)\\ 
$5$ \hphantom{00} & 9 (12.4$\pm$4.2) & 11 (9.0$\pm$5.0) &    11 (12.4$\pm$4.2) & 12 (9.2$\pm$4.9)\\
$6$ \hphantom{00} & 17 (10.5$\pm$3.9) & 8 (7.6$\pm$4.6) &  15 (10.5$\pm$3.9) & 7 (7.8$\pm$4.6)\\
$7$ \hphantom{00} & 6 (9.1$\pm$3.7) & 6 (6.6$\pm$4.3) &  7 (9.1$\pm$3.7) & 6 (6.7$\pm$4.3) \\
$8$ \hphantom{00} & 8 (8.0$\pm$3.5) & 4  (5.8$\pm$4.0) &  7 (8.0$\pm$3.5) & 4 (5.9$\pm$4.0) \\
$9$ \hphantom{00} & 8 (7.2$\pm$3.3) & 5 (5.2$\pm$3.8)    &  8  (7.2$\pm$3.3) & 5 (5.3$\pm$3.8) \\ 
\botrule
\textbf{Pearson} $\chi^{2}$ \hphantom{00} & \bf10.375 &  \bf6.178 & \bf7.028  & \bf3.932 \\ \botrule
\textbf{K-S test}  \hphantom{00} & \bf0.048$<$0.109 &  \bf0.047$<$0.127 & \bf0.048$<$0.109  & \bf0.032$<$0.126 \\ \botrule
\end{tabular} \label{ta3}}
\end{table}

\begin{table}[h]
\tbl{The distribution of leading digits of worldwide IFFs}
{\begin{tabular}{@{}lllllllllll@{}} \toprule
Sample \hphantom{00} & Size & \textbf{Pearson} $\chi^{2}$ & \textbf{K-S test} \\\botrule
All Normalized \hphantom{00} & 659 & 9.362 & 0.026$<$0.053\\
All Non-normalized \hphantom{00} & 928 & 13.838 & 0.029$<$0.045\\ 
All IFFs estimates \hphantom{00} & 1587 & 13.272 & 0.022$<$0.034\\\botrule
\end{tabular} \label{ta1}}
\end{table} 
In Table 3 we present the analysis for IFFs from least developed countries\cite{kar}. As seen from eq. (3) the width of the CI is inversely proportional to the size of the sample. Thus for small sample sizes the CI tend to be wide and produce less precise results. For the least developed countries instead of CI we give the root mean square error ($\Delta{N}$) for each digit calculated from the binomial distribution\cite{Shao}
\begin{equation}
\Delta{N}= \sqrt{NP(d)(1-P(d))}
\end{equation} 
The $\chi^{2}$ and K-S statistic (last two rows and column 2-4) are all less than their respective critical values and therefore null hypothesis is accepted. Next we turn our attention to the January 2011 report of GFI which gives the estimates of the IFFs for the period 2000-2008\cite{kar2}. The statistical analysis of (Tables 3, 4 and 7) this report is shown in Table 4. Finally in Table 5 we show the analysis for the IFFs estimates (Tables 4, 5 and 9) of the December 2011 report of GFI. As shown the $\chi^{2}$ for all the four columns are less than the critical value of 15.507. Similarly for each column the K-S test statistic are less than the corresponding critical values. A graphical representation of the results obtained in Tables 2-5 is given in Figs. 1-4 which clearly demonstrate the  occurrence of the significant digits for IFFs data as predicted by Benford's law.
\begin{figure}
\begin{minipage}[b]{.9\linewidth}
\vspace*{-5pt}
\hspace*{5pt}
\centering
\begin{tabular}{cc}
\hspace*{-65pt}
\vspace*{-90pt}
\epsfig{file=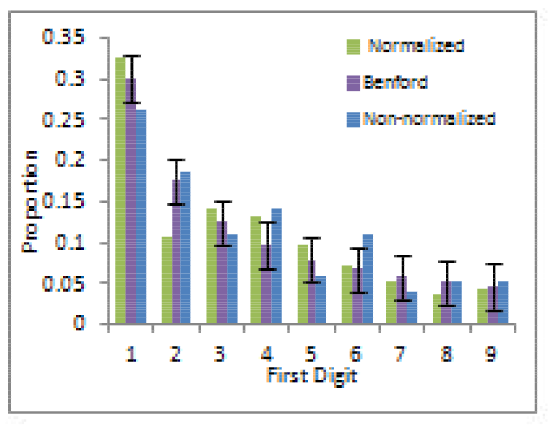,width=0.7\linewidth, height=0.7\linewidth, angle=270, clip=}&
\hspace*{-60pt}
\epsfig{file=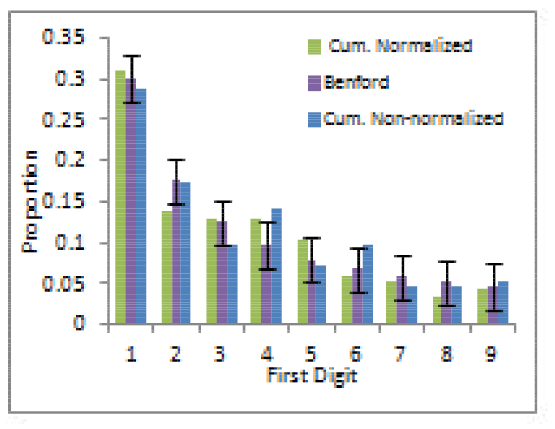,width=0.7\linewidth, height=0.7\linewidth, angle=270,  clip=} \\
\hspace*{-40pt}
\vspace*{40pt}
\hspace*{-130pt}
\end{tabular}
\vspace*{35pt}
\hspace*{20pt}
\end{minipage}
\caption{Observed and Benford distributions of significant digits for country-wise largest average normalized, largest average non-normalized, cumulative normalized and cumulative non-normalized IFFs: 2000-2009 (millions of U.S. dollars)}
\end{figure}
\begin{figure}
\begin{minipage}[b]{.9\linewidth}
\vspace*{-5pt}
\hspace*{5pt}
\centering
\begin{tabular}{cc}
\hspace*{-65pt}
\vspace*{-90pt}
\epsfig{file=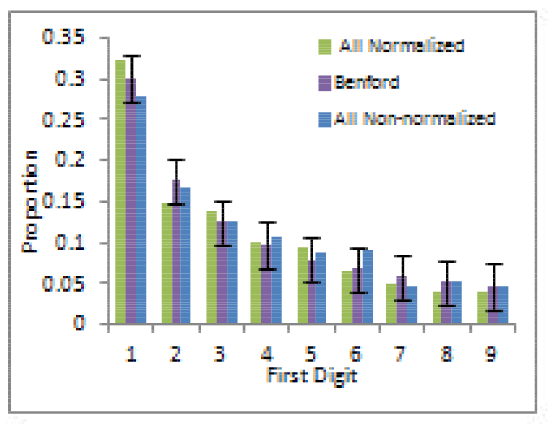,width=0.7\linewidth, height=0.7\linewidth, angle=270, clip=}&
\hspace*{-60pt}
\epsfig{file=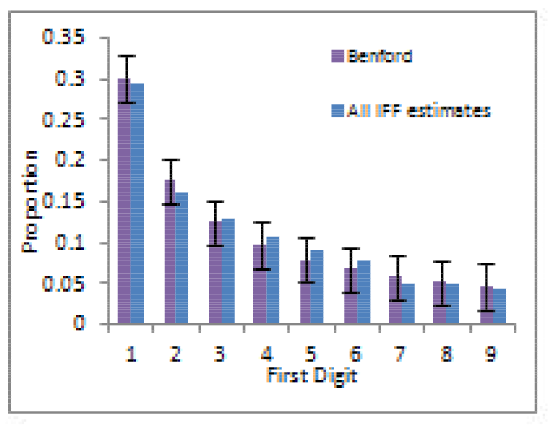,width=0.7\linewidth, height=0.7\linewidth, angle=270,  clip=} \\
\hspace*{-40pt}
\vspace*{40pt}
\hspace*{-130pt}
\end{tabular}
\vspace*{30pt}
\hspace*{20pt}
\end{minipage}
\caption{Observed and Benford distributions of significant digits for country-wise all normalized, all non-normalized and all IFFs estimates (millions of U.S. dollars)}
\end{figure}
\newline
The agreement of the observed and the predicted digit distributions in each group studied is impressive. However, being conscious of the fact that as the number of records increases the occurrence of first digits in a given data set tends to be more closer to a Benford distribution\cite{Nigrini3}. Therefore, we first form two large data sets by combining separately the normalized and the non-normalized IFFs estimates and then all the estimates into a further larger data set. The relevant samples, size and the corresponding Pearson's $\chi^{2}$ and K-S test statistics are given in Table 6. We also show that critical values of K-S test statistic for each of the samples. It can be seen that the $\chi^{2}$'s and K-S test statistic are less than their critical values indicating a clear agreement between the theoretical and the observed distributions of the leading digits as is illustrated in Fig. 5. 

\section{Discussion}
The volume of IFFs from developing countries far exceeds the assistance they receive from their developed counterparts making the former net creditor to the latter a fact which is quite contrary to the common perception of the reverse being true\cite{Ndikumana}. Further, preventing these economic resources to work and generate revenues in a country's legal framework, IFFs adversely impact the economic growth and in many cases play a significant role in the financial crisis of countries\cite{Fabre}. The inter-linkage of different economies in present global structure in turn may set off a chain reaction leading to financial crisis on a bigger scale e.g. the current financial crisis in Eurozone. Thus, in the quest of reforming the global financial system the IFFs have emerged as one of the central issues of international and national policy agendas\cite{tax, Reuter}. 
The interest generated by the GFI reports has however also triggered a debate on the validity of the estimates of IFFs\cite{Reuter}. To contribute to this debate we performed a statistical assessment of these estimates using a well known mathematical regularity on the distribution of first digits called Benford's law. 
\newline 
Data in which the occurrence of numbers is free from restrictions have a tendency to follow Benford's law. Manipulated, invented and influenced numbers do not follow the law \cite{Hill2}. However, Benford's law is not definitive\cite{Judge} in the sense a deviation from it does not prove manipulation just as the conformity to it does not prove the truthfullness and further research is necessary before making any final decision on the quality of the data. Nevertheless Benford's law is useful in analytical procedures for testing the completeness of financial reports\cite{Nigrini2, Bhattacharya}.
\newline
Benford's law has been proved to successfully unravel falsification of financial documents, tax evasion by individuals, manipulated trade invoices and tax returns submitted by the companies\cite{Nigrini1, Kumar}, the illicit practices which significantly contribute to the IFFs\cite{Ndikumana}.
We have used $\chi^{2}$ and K-S test to assess the overall significance of departures of the observed frequencies of the digits and those expected on the basis of Benford's law and found the difference between actual and theoretical frequencies to be insignificant at the 5\% level of significance which means that IFF data follow the law.
\newline
Normalized data sets in general have more significant (smaller) $\chi^{2}$ than corresponding non-normalized ones indicating their better submission to Benford's law. This can be readily appreciated from figures where digit proportions of the normalized and corresponding non-normalized data sets of each report along with Benford proportion have been compared. The improvement in $\chi^{2}$ may be due to the reduced number of records in normalized data sets. The normalization just reduces the number of countries in the data without affecting any change in occurrence of the first digits.  However, a manual scan of the data also revealed appreciable decrease in the volume of IFFs which in turn lead to change in initial digits for several countries. For example, for the period of 2000-2009\cite{kar2}, for Russia the non-normalized and normalized values in millions of US dollars are 53141 and 47,478 respectively. For Ukraine, the respective IFFs for the same period are 10,757 and 9152. Similarly, for Brazil the non-normalized estimate is 7317 whereas the normalized estimate is 2614. Further, for a transformation to be free from any flaws it must preserve any Benford like nature of the input data i.e. post transformation data must also be Benford like\cite{Varian}. Thus submission of both the non-normalized and the normalized IFF estimates, input and output for normalization, to Benford's law might be an indication of validity of the normalization process.
\newline
The current financial crisis has severely reduced the capacity  of the developed countries to provide assistance to the developing countries. Therefore, developing countries, have been forced to uncover their own financial resources and thus guided by the GFI estimates of IFFs, have initiated measures for curbing the exit of illicit financial flows besides recovering their stolen assets. An extensive review on comparison of various estimates of IFFs has found that those of GFI correspond well with the calculations of other researchers\cite{tax}. Particularly, for the least developed countries, a majority of which belong to African continent, the studies \cite{kar, Ndikumana} found similar estimates of capital flight in spite of the difference in sample size and considerable data issues. 
\section{Conclusion}
IFFs from developing countries have generated a lot of public interest. We analyzed the distribution of leading digits of data on the IFFs from developing countries and found that data follows the predictions of the Benford's law. In the light of the present global financial crisis and the debate it has triggered on the illegal capital flows of the countries, we think our study may further contribute to the understanding of the IFFs estimates. 
 \section*{Acknowledgments}
The author thanks GFI for free access to data and Dev Kar for helpful comments. Suggestions from P. M. Ishtiaq are gratefully acknowledged. 



\end{document}